\documentclass[prd,twocolumn,amsmath,amssymb]{revtex4}

\usepackage{amsmath}
\usepackage{amssymb}
\usepackage{graphicx}

\bibliography{ref}
\usepackage{float}

\usepackage[T1]{fontenc} 
\usepackage{indentfirst}
\begin{document} 
\title{Magnetic and Electronic Structure of the Film-Stablized Mott Insulator  BaCrO$_3$ } 
\author{Z. H. Zhu,$^1$ F. J. Rueckert,$^1$ J. I. Budnick,$^{1, 2}$ W. A. Hines,$^1$ M. Jain,$^{1, 2}$  H. Zhang,$^2$ and B. O. Wells$^{1, 2}$\\ $^1$Department of Physics, University of Connecticut, Storrs, Connecticut 06269-3046, USA\\ $^2$Institute of Material Science, University of Connecticut, Storrs, Connecticut 06269-3136, USA} 
\date{\today}

\begin{abstract} We have synthesized and characterized laser-deposited film samples of perovskite BaCrO$_3$, a missing member of the perovskite-chromate family. The BaCrO$_3$ films have a substantially larger lattice constant than other chromates, are insulating, and exhibit weak ferromagnetism. Employing first-principles density functional theory calculations, we attribute such weak ferromagnetism to a canted C-type antiferromagnetic spin order. Additionally, comparison with the other sister compounds CaCrO$_3$ and SrCrO$_3$ suggests an anomalous Mott transition where magnetism is independent of whether the compound is metallic or insulating.
\end{abstract}
\maketitle

Transition metal oxides (TMOs) with perovskite and related structures exhibit many fascinating electronic phenomena, including Mott insulators, metal-insulator transitions, and high-Tc superconductivity, due to transport being determined by strongly correlated d electrons.$^1$ One of the ubiquitous questions for these compounds is how exactly transport is coupled to magnetic degrees of freedom.  The textbook situation is that ferromagnetism leads to metallic conductivity, while antiferromagnetism is associated with non-conducting behavior, particularly Mott insulators. A classic example is colossal magnetoresistance in the manganites, where a magnetic field can induce a transition from an insulating antiferromagnetic phase to a conducting ferromagnetic phase.$^2$ There are some exceptions which indicate other new phenomena may be in play. Observations of ferromagnetic insulators and antiferromagnetic conductors are most often explained by particular orbital orderings$^3$ or reduced electronic and structural dimensionality.$^4$ This report describes the properties of a new end member of a family that can be driven through a metal-insulator transition using chemical pressure, or bandwidth control, yet seems to retain the same magnetic structure on both sides of the transition. We have synthesized BaCrO$_3$ in perovskite form by  using epitaxial film growth on an appropriate crystal template.\vspace*{0.1cm} This polymorph is apparently not stable in the bulk.

Four decades ago, a few chromate perovskites ACrO$_3$ (A = Ca, Sr, and Pb) were synthesized by several groups using solid-state reaction techniques with high pressure ($\approx$ 6-10 GPa) and high temperature ($\approx$ 1000 K).$^{5-9}$  Recent studies on  CaCrO$_3$ and SrCrO$_3$ have exhibted some novel properties, drawing more attention to these chromate compounds.$^{10-16}$ For example, both of these compounds are antiferromagnetic conductors. It is also suggested that CaCrO$_3$ possibly lies in the crossover region between being itinerant and localized, implying the necessity to consider correlation effects.$^{14, 15}$ An interesting orbital ordering transition and electronic phase coexistence have been discovered in SrCrO$_3$.$^{12}$Among these chromate compounds, only PbCrO$_3$ is unambiguously thought to be an insulator with the G-type AFM spin structure. However, because of the different outer orbital configuration of Pb, PbCrO$_3$ can hardly be used to form a unified scheme to understand the exotic properties of the other chromate family compounds. On the other hand, perovskite BaCrO$_3$, which should be akin to CaCrO$_3$ and SrCrO$_3$, has not been formed as yet using traditional solid-state reaction methods. It seems that BaCrO$_3$ tends to form a hexagonal phase.$^{17,18}$ This can be understood by considering the tolerance factor t = (r$_A$+r$_O$)/[$\sqrt{2}$(r$_B$+r$_O$)], where r$_A$, r$_B$, and r$_O$ are the ionic radii of the A, B cations, and oxygen anion, respectively.$^{19}$ Typically, perovskite structures have t < 1, while hexagonal polytypes result when t > 1. The tolerance factor of BaCrO$_3$ is 1.031, which is slightly above the allowed range for a cubic perovskite phase.  Epitaxial film growth on an appropriate substrate has been known as a method for stabilizing particular phases. Here we have successfully used well lattice-matched SrTiO$_3$(STO) (001) surfaces as a template for growing perovskite BaCrO$_3$ using pulsed laser deposition (PLD) and are able to measure its basic properties. 

	The laser target was a stoichiometric mixture of BaO and Cr$_2$O$_3$ prepared by solid-state reaction at 1000 $^\circ$C in air.  During growth, the substrate temperature was kept at 800 $^\circ$C in 10$^{-7}$ torr vacuum. After deposition, the films were cooled to room temperature at a rate of 4$^\circ$C per minute. The film growth was found to be extremely sensitive to both the vacuum conditions and temperature of the substrate. Typically, to grow oxides one introduces oxygen into the chamber during the growth process. We found that in a large temperature range (mostly below 700 $^\circ$C) for most any oxygen partial pressure that the film favors the Hashemite BaCrO$_4$ structure. Under the assumption that too much oxygen might be a problem, the films then were grown under a vacuum condition and various temperatures. Eventually, we were able to synthesize the perovskite BaCrO$_3$ films at 800  $^\circ$C in 10$^{-7}$ torr vacuum.

The crystal structure of the BaCrO$_3$ film was determined using both a two-circle x-ray diffractometer to study peaks perpendicular to the surface and a three-circle diffractometer with an area detector for determining in- and out-of-plane lattice constants. All of the x-ray data used Cu K$\alpha$ ($\lambda$ = 1.5406\AA) radiation. X-ray photoemission spectroscopy (XPS) was performed in order to study the composition and purity of the BaCrO$_3$ films using a PHI Multiprobe with Al anode.  Measurements of the dc magnetization were made on a Quantum Design MPMS SQUID magnetometer for temperatures 5.0 K $\leq$ T$ \leq$  300 K and magnetic fields -50 kOe $\leq$ H $\leq$ +50 kOe. Resistivity measurements were carried out using a conventional four-probe technique along with the variable temperature and magnetic field control of the MPMS. Finally, first principle calculations of the magnetic order and electronic structure of constraint representative 2x2x2 supercells of BaCrO$_3$ was carried out using the Vienna \emph{ab} initio simulation package (VASP) $^{20-22}$. More details of calculation can be seen in Ref. 23.
\begin{figure}[htb]
\centering
\includegraphics[scale=0.3]{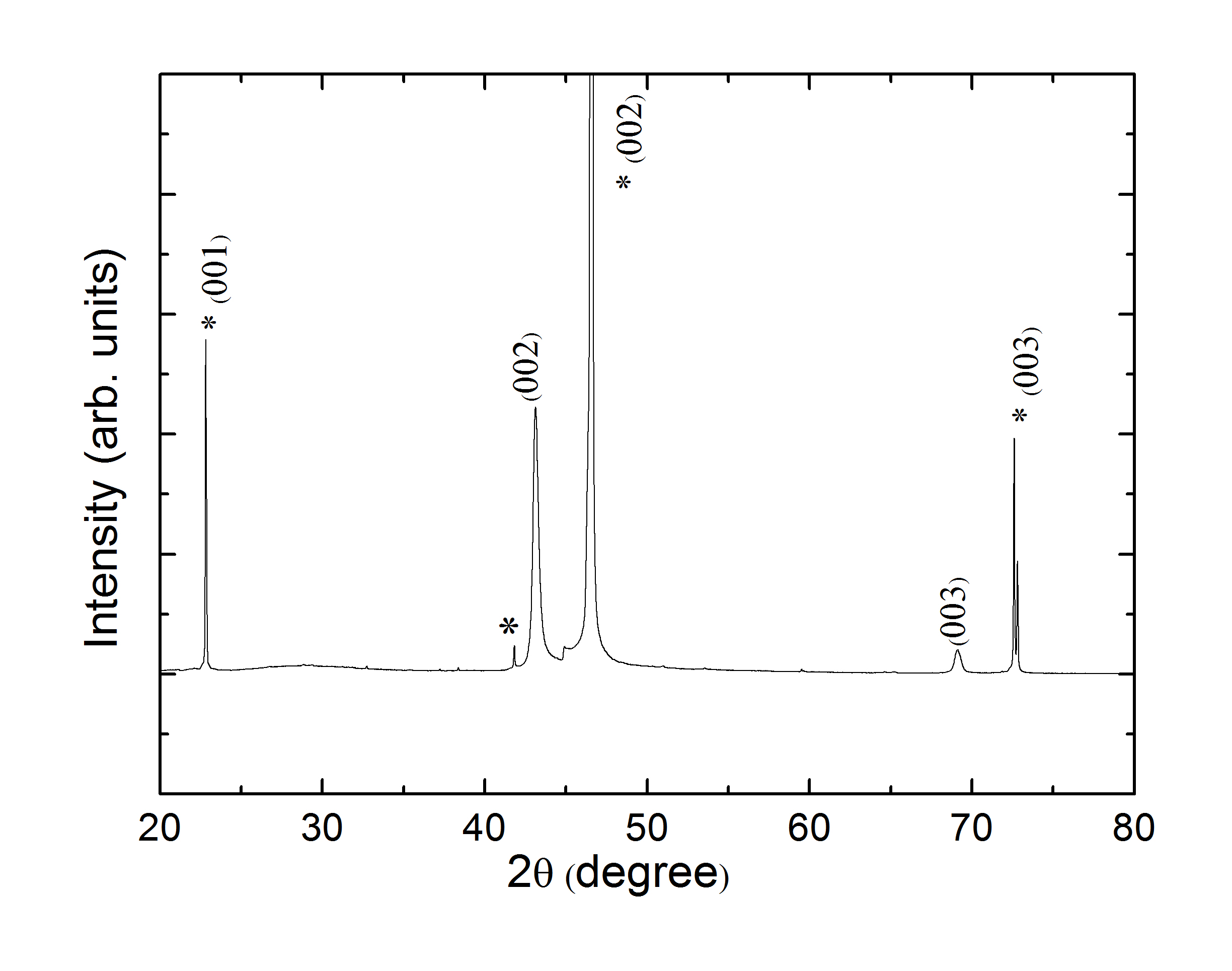}
\caption{\label{fig:epsart}X-ray diffraction pattern for a typical epitaxial BaCrO$_3$ thin film on a STO(001) substrate.  The (002) and (003) BaCrO$_3$ film peaks, which occur at 43.12$^\circ$ and 69.09$^\circ$, respectively, identify the $``$cubic $"$ perovskite . The small peak labeled with (*) at 41.82$^\circ$ corresponds to the STO substrate (002) K$_\beta$ reflection. The small step-like feature at 44.90$^\circ$ is an experimental artifact related to the window size of the detector.}
\vspace{-0.50cm}
\end{figure}

Figure 1 shows the x-ray diffraction (XRD) pattern for out-of-plane peaks of a typical epitaxial BaCrO$_3$ film on a STO(001) substrate. The BaCrO$_3$ film peaks at 43.12$^\circ$ and 69.09$^\circ$ correspond to the (002) and (003) reflections, respectively, and identify the $``$cubic$"$perovskite structure.  This result is unlike the case for powders where the hexagonal structure has been observed.$^{17,18}$ The lattice constants are measured on an oxford single crystal diffractometer and give the following  room-temperature values: out-of-plane lattice constant c = 4.07 \AA , in-plane lattice constant a = 4.09 \AA. In comparison, the pseudocubic lattice parameter for bulk CaCrO$_3$ and SrCrO$_3$ are 3.75 \AA  and 3.82 \AA, respectively. The trend is that as the atomic number increases from Ca (20) to Ba (58), the lattice constant increases significantly. The structure of our film samples thus has a small tetragonal distortion from the cubic case. Tilting patterns in the BO$_6$ octahedra are common in perovskites, although in this case the associated superlattice peaks appear to be too weak to be detected from our film samples.

\begin{figure}[htb]
\includegraphics[scale=0.30]{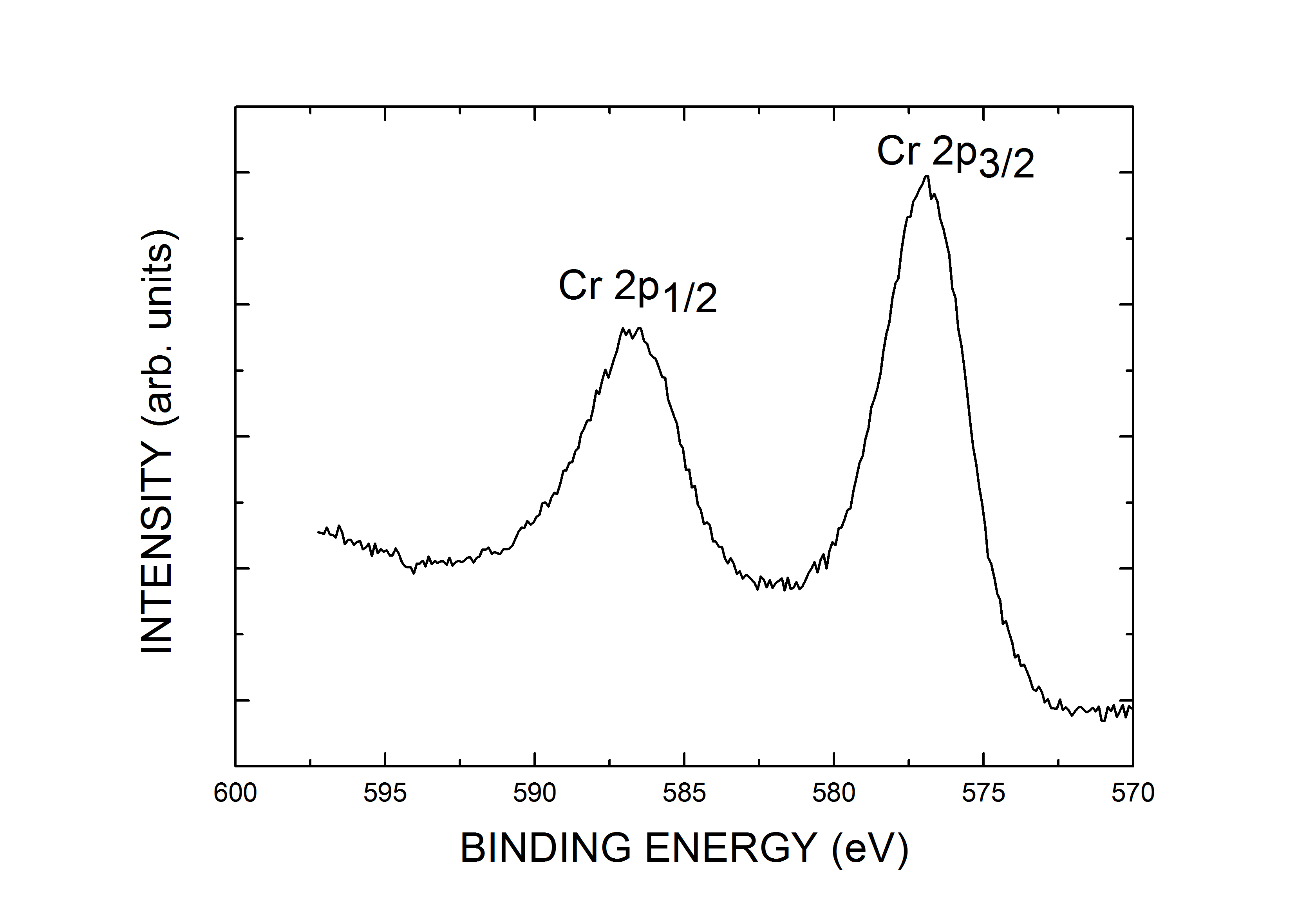}
\caption{\label{fig:epsart} X-ray photoemission spectrum for a typical epitaxial BaCrO$_3$ thin film on a STO(001) substrate.  The binding energies for the Cr core level 2p$_{3/2}$ and 2p$_{1/2}$ peaks are 577.1 eV and 586.8 eV, respectively, consistent with the Cr$^{4+}$ valence state for the BaCrO$_3$ film.}
\vspace{-0.25cm}
\end{figure}

X-ray photoemission spectroscopy (XPS) was performed in order to study the chemical composition and phase purity of the film, along with the ionic state of Cr. No extra peaks due to impurities were observed except for that of Au, which is used for calibration and grounding of the film.  The ratio between the Ba and Cr is found to be 1:1 and remains constant throughout the thickness of the film, which was revealed by collecting spectra as we sputtered through the film.  A scan of the Cr 2p core level region is shown in Fig. 2.  The binding energies for the 2p$_{3/2}$ and 2p$_{1/2}$ peaks of Cr are 577.1 eV and 586.8 eV, respectively, which matches binding energies of main peaks due to spin-orbit splitting for CaCrO$_3$, indicating the Cr$^{4+}$ valence state in the BaCrO$_3$ films.$^{16}$ Our spectrum in Fig. 2 overall is very similar to that for CaCrO$_3$ in Ref. 16, athough we appear to have slightly poorer resolution, which does not allow us to detect the additional shoulder-like features reported in the latter

The BaCrO$_3$ films have a small, but definite magnetic response. Measurement of this signal is complicated by the small size of the film signal compared to the relatively large contribution of the substrate.  Magnetic measurements were carried out for the magnetic field aligned both parallel and perpendicular to the film/substrate.  For the parallel case, the samples were mounted onto a uniform quartz rod with a very small amount of two-sided tape.  For the perpendicular case, the samples were inside a small diameter plastic straw, held in place by cotton.  The magnetic characteristics of a blank STO(001) substrate were measured with both orientations for temperatures 5 K $\leq$ T $\leq$ 300 K and magnetic fields -50 kOe $\leq$ H $\leq$ +50 kOe.  In addition to the diamagnetic response of the STO substrate material, other background effects were present.  These included: (1) trace amounts of magnetic impurities in the STO substrate material, (2) magnetic response of the cotton, which was predominately diamagnetic, and (3) trace amounts of oxygen in the SQUID sample chamber. The presence of oxygen, which can be significantly reduced but not always completely eliminated, frequently complicates low temperature magnetic measurements.  Nevertheless, the various background effects were accounted for in order to isolate the magnetic response of the BaCrO$_3$ film.  Fig. 3(a) shows the zero-field-cooled (ZFC) and field-cooled (FC) curves obtained for a magnetic field H = 500 Oe (H parallel) from a typical BaCrO$_3$ film on a STO(001) substrate. In order to account for the background contributions, corresponding to ZFC and FC curves at H = 500 Oe were obtained for a blank STO(001) substrate (see Fig. 3(b) ). The bifurcation of the ZFC and FC curves that occurs at T $\approx$ 200 K for the STO substrate is attributed to magnetic impurities in the substrate material, while the small feature at T $\approx$ 48 K is due to trace amounts of residual oxygen in the sample chamber. Although not perfect, subtracting the subtrate curves from the corresponding total moment curves to yield the BaCrO$_3$ film curves [i.e., 3(a) - 3(b) = 3(c)] nearly removes these contributions. It can be seen in Fig. 3(c), that there is a peak in the ZFC curve for the BaCrO$_3$ thin film at T $\approx $25 K, below which the ZFC and FC show a significantly different temperature dependence.  As discussed below, this behavior was always reproducible and is attributed to a magnetic transition associated with the BaCrO$_3$ film.  

\begin{figure}
\includegraphics[scale=0.3]{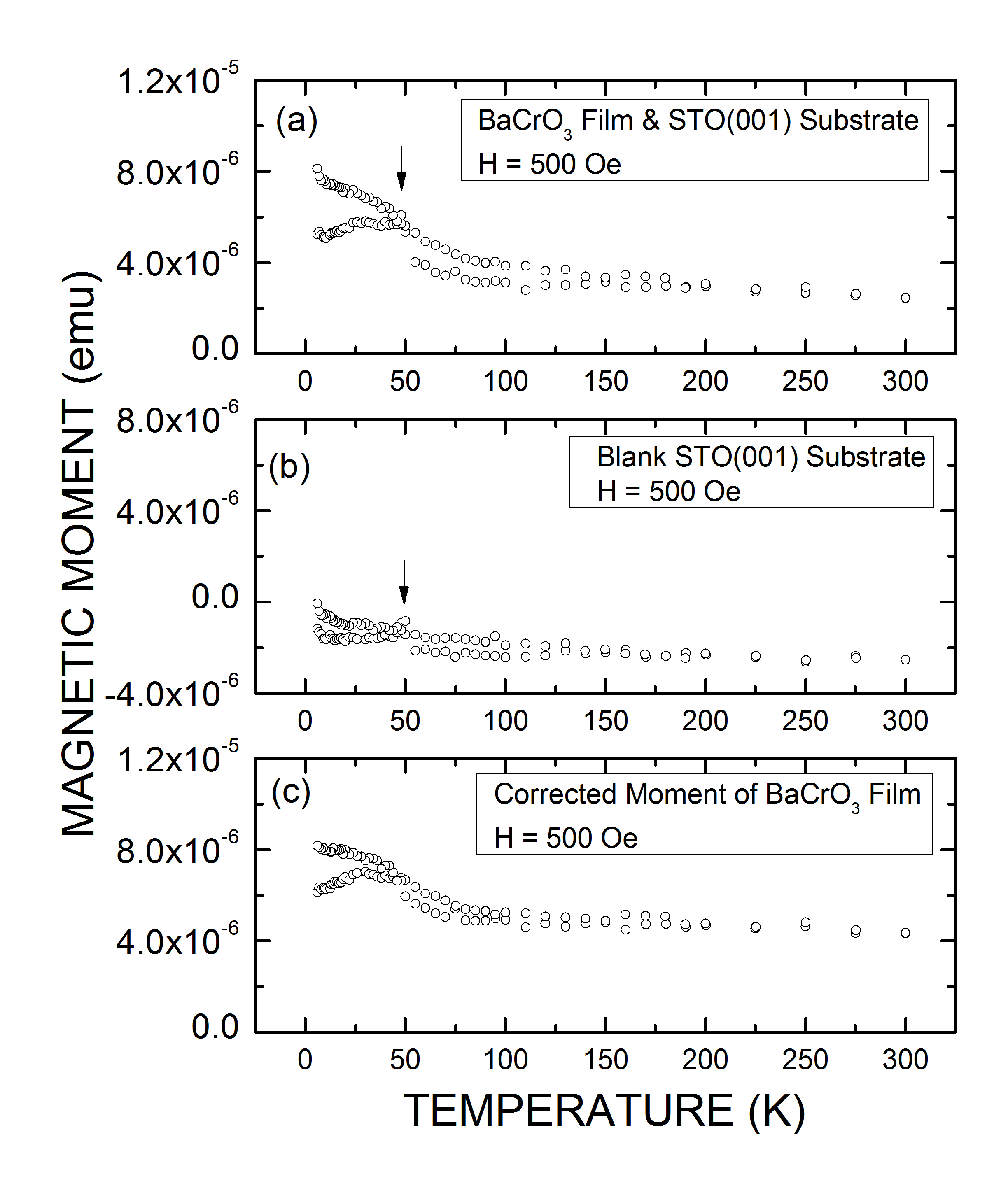}
\caption{\label{fig:epsart}Zero-field-cooled (ZFC) and field-cooled (FC) dc magnetization versus temperature for a magnetic field H = 500 Oe (H parallel) from a typical: (a) epitaxial BaCrO$_3$ thin film on a STO(001) substrate, (b)  blank STO(001) substrate, and (c) epitaxial BaCrO$_3$ thin film only, obtained  by subtraction. There is a peak in the ZFC curve at T $\approx$ 25 K, below which the ZFC and FC curves show a significant difference in temperature dependence (see Fig. 3(c)).  This behavior is attributed solely to the BaCrO$_3$ film and suggests the onset of weak ferromagnetism. The bifurcation of the ZFC and FC curves that occurs at T $\approx$ 200 K is attributed to magnetic impurities in the substrate material, while the small feature at T $\approx$ 48 K is due to a trace amount of residual oxygen in the sample chamber.}
\end{figure}

\begin{figure}[htb]
\includegraphics[scale=0.25]{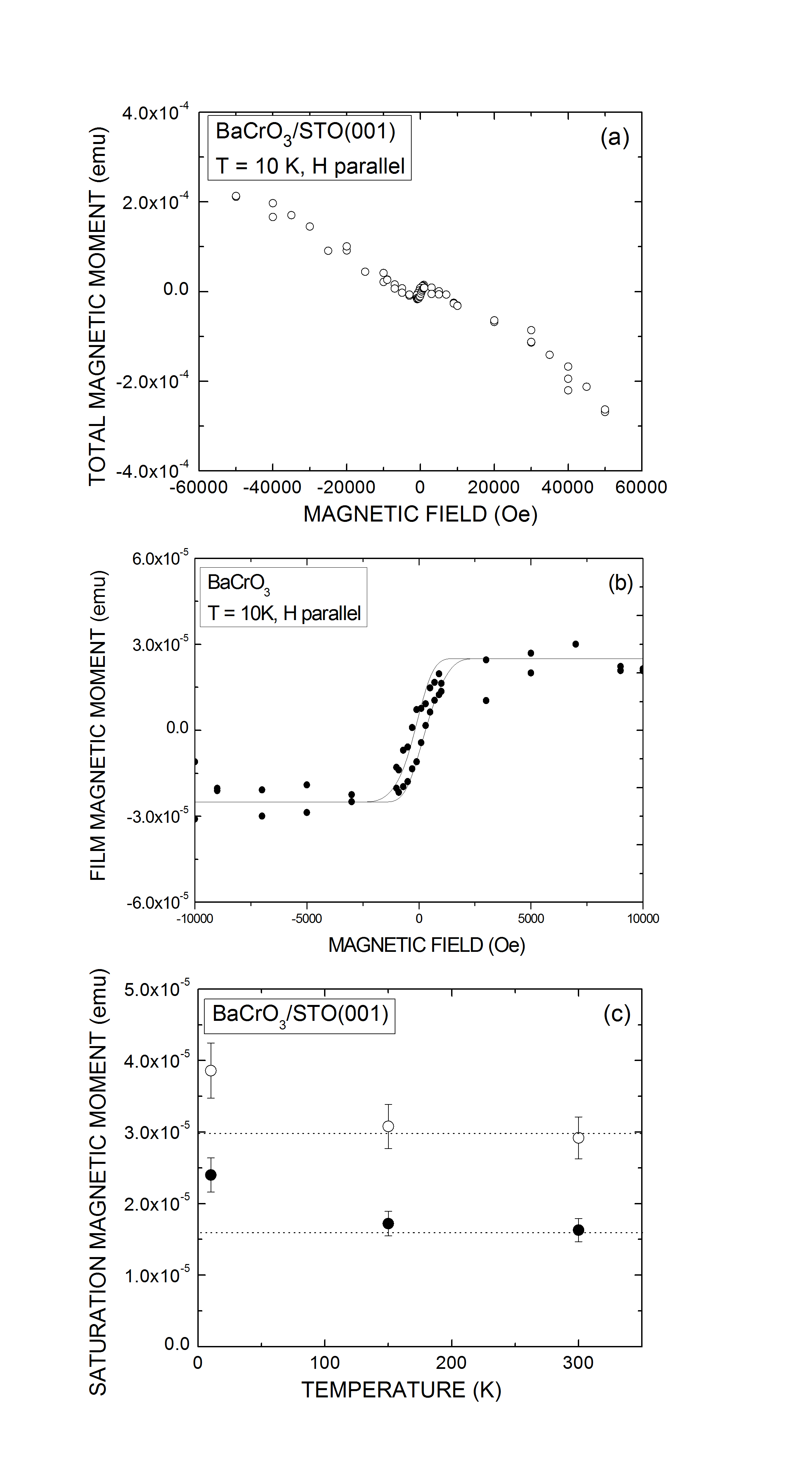}
\caption{\label{fig:epsart}(a) Hysteresis loop magnetization at 10 K for a typical BaCrO$_3$ thin film on a STO(001) substrate with H parallel to the film/substrate.  The negative slope reflects the dominant diamagnetic response for the STO(001) substrate.  (b) Hysteresis loop magnetization from (a) corrected for the diamagnetic contribution of the substrate.  The resulting curve is characteristic of weak ferromagnetic behavior with a small coercive field and a magnetization that saturates for H $\geq$ 1,000 Oe. The solid line is a guide to the eye. (c) Saturation magnetic moment values versus temperature obtained from curves such as shown in (b).  The closed (open) symbols represent values obtained for H parallel (perpendicular) to the film/substrate.  The moment values obtained for T = 150 K and 300 K (above the 25 K transition temperature) are the background contributions from the substrate and cotton, while the larger moment values obtained at T = 10 K include the contribution from the BaCrO$_3$ film.}
\vspace{-0.80cm}
\end{figure}

\begin{figure}[htb]
\includegraphics[scale=0.25]{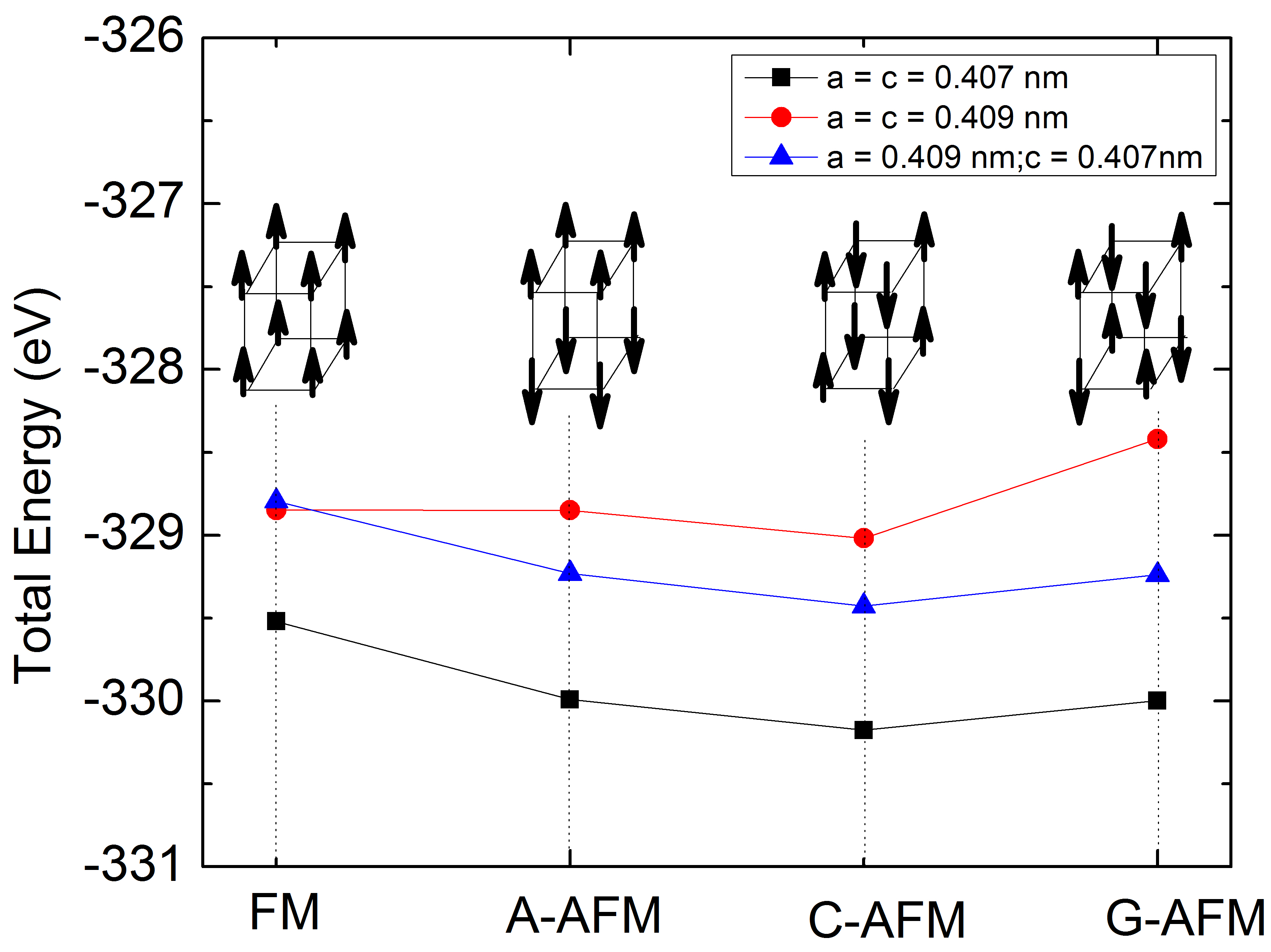}
\caption{\label{fig:epsart} Calculated total ground state energies of various spin orders in three typical lattice strcutures for BaCrO$_3$ based on experimental observations.}
\vspace{-0.2cm}
\end{figure}
 In order to explore the nature of the magnetic ordering for the BaCrO$_3$ film on a STO(001) substrate, full hysteresis loops (-50 kOe $\leq$ H $\leq$ +50 kOe) were obtained for temperatures above and below the T $\approx$ 25 K transition, principally for T = 300 K, 150 K, and 10 K.  All of the loops obtained show a dominant diamagnetic contribution due to the STO(001) substrate along with a much smaller ferromagnetic contribution.  As an example, Fig. 4(a) shows the total as-measured hysteresis loop magnetization obtained at 10 K from a sample consisting of a typical BaCrO$_3$ film on a STO(001) substrate with H parallel to the film/substrate.  Fig. 4(b) shows the hysteresis loop magnetization in Fig. 4(a) with the diamagnetic part of the substrate contribution subtracted off.  The resulting curve in Fig. 4(b) is characteristic of weak ferromagnetic behavior with a small coercive field and a magnetization that saturates for H $\geq$ 1,000 Oe.  Fig. 4(c) shows the values of the saturation magnetic moment, for T = 10 K, 150 K, and 300 K, obtained from curves such as shown in Fig. 4(b).  The closed (open) symbols represent values obtained for H parallel (perpendicular) to the film/substrate.  The saturation magnetic moment values obtained for T = 150 K and 300 K (above the 25 K transition temperature) are the background contributions from the substrate and cotton.  Similar measurements on blank substrates revealed that the background contributions were essentially temperature independent.  On the other hand, the saturation magnetization values obtained at T = 10 K (below the 25 K transition temperature) shown in Fig. 4(c) include a contribution from the BaCrO$_3$ film.  This behavior was reproducible in other film/substrate samples.  Thus, the saturation moment for epitaxial BaCrO$_3$ film can be obtained by subtraction.  Taking the saturation magnetic moment value of 0.7 $\times$ 10$^{-5}$ emu (see Fig. 4(c)), with a film thickness $\approx$ 100 nm, the magnitude of the magnetic moment per Cr is estimated to be 0.028 $\mu$$_B$, too small to attribute to fully ferromagnetically-aligned localized Cr$^{4+}$ ions (2 $\mu$$_B$/Cr). 
Instead, similar to CaCrO$_3$, it seems reasonable to attribute such weak ferromagnetism to a canted antiferromagnetic spin structure.$^{10,24}$  The rotational distortion of the CrO$_6$ octahedra that likely causes the measured tetragonal phase in this sample also allows for a Dzyaloshinsky-Moriya (DM)$^{25,26}$ interaction, similar to the case for the canted Cu$^{2+}$ moments in La$_2$CuO$_4$. To explore the expected magnetic structure in more detail, we performed $\emph {ab}$ initio electronic structure calculations fixing the lattice to the measured structure, along with some possible variations. We explore the following magnetic configurations: ferromagnetic, A-type antiferromagnetic, C-type antiferromagnetic, and G-type antiferromagnetic. As shown in Fig. 5, for all of the structure variations explored the most favorable magnetic structure is C-type antiferromagnetism. This matches the observed and calculated magnetic structure of conducting CaCrO$_3$.$^{13}$ The calculated magnetic moment $\mu$ = 1.70 $\mu$$_B$/Cr for BaCrO$_3$  is greater than that of CaCrO$_3$ (1.52 $\mu$$_B$/Cr calculated, 1.20 $\mu$$_B$/Cr experimental), indicating a more localized behavior of Cr$^{4+}$ in BaCrO$_3$, naturally associated with its larger lattice constant weakening the \emph{pd} hybridization. Further measurements with a direct probe such as neutron diffraction would be necessary to definitely determine the magnetic structure. Our assignment of a canted C-type antiferromagnetic structure is consistent with magnetization measurements, first principle calculations, and the related compound CaCrO$_3$, and is the best estimate practical for the thin film samples.

  \begin{figure*}
\includegraphics[scale=0.65]{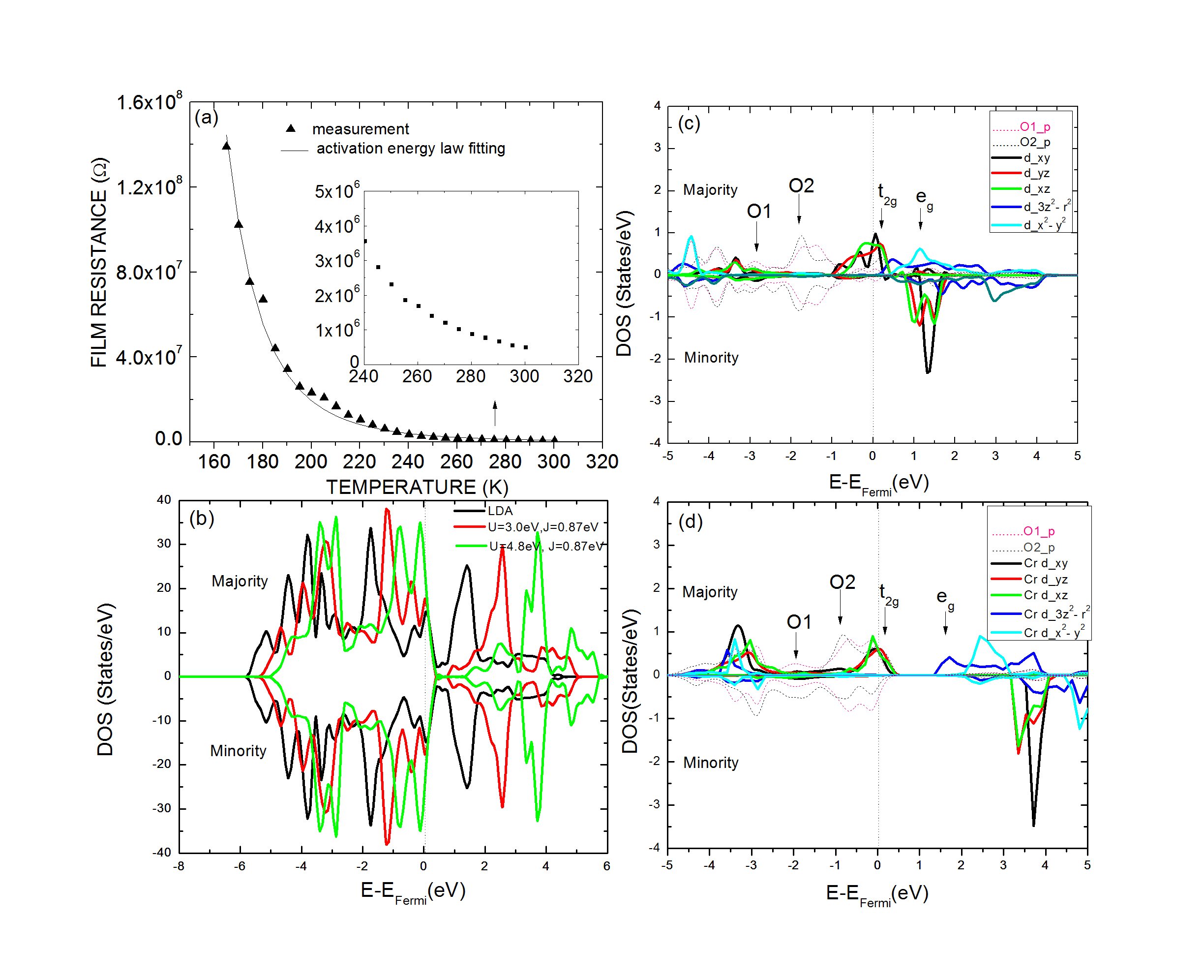}
\caption{\label{fig:wide}(a) Resistance versus temperature for a typical epitaxial BaCrO$_3$ thin film on a STO(001) substrate: closed triangles -measured values and solid line - fit to activation law yielding an energy parameter $\Delta$ $\approx$  0.38 eV. (b) Total density of states (TDOS) of BaCrO$_3$ with room temperature crystal structure and C-type AFM spin order using LSDA and LSDA+U. (c) and (d) Projected density of states (PDOS) of  the d orbital of Cr atom  and the  p orbital of O atom calculated by LSDA and LSDA+ U (U = 4.8 eV, J = 0.87 eV) in C-AFM order, where O1 and O2 represent in-plane and apical O atoms, respectively.   }
\vspace{-0.1cm}
\end{figure*}
Fig. 6(a) shows the measured BaCrO$_3$ film resistance (closed triangles) as a function of temperature revealing that BaCrO$_3$ is insulating.  The variation of the resistance with temperature can be fit with the activation law $\rho$ = $\rho$$_0$ exp[ -$\Delta$/(k$_B$T)] yielding $\Delta$ $\approx$ 0.38 eV, which is smaller than that previously reported for PbCrO$_3$ ($\approx$ 0.5 eV).$^{14}$  However, our electronic calculation for C-type AFM BaCrO$_3$ using a spin polarized local density approximation (LSDA) as shown in Fig. 6(b) exhibits a metallic behavior. Conductivity is more difficult to describe by $ \emph {ab}$ initio methods than ground state magnetic structure but when the primary difficulty is the strength of electronic correlations in compounds with partially filled d bands an LSDA+U calculation may capture the basic electronic state. For these calculations, we chose J = 0.87 eV and U = 3 eV as calculated for CrO$_2$ and CaCrO$_3$, respectively.$^{13, 27}$ We included another calculation with a larger U = 4.8 eV based on a recent experimental report on CaCrO$_3$ using resonant photoemission.$^{16}$ As shown in Fig. 6(b) the ground state remains  metallic within coulomb corrections, which conflicts with our experimental observations. To further elucidate the emergence of a gap opening above the Fermi surface after  introducing a coulomb correction, projected density of states (PDOS) of the d orbital of Cr atom and the p orbital of O atom are calculated by using LSDA and LSDA+U, shown in Fig. 6(c ) and Fig. 6(d),  respectively.  It can be seen that the gap opening stems from the splitting between the t$_{2g}$ and the e$_g$ orbitals of the Cr due to the coulomb interaction.  However, the Fermi energy still lies in the t$_{2g}$ orbitals of Cr and p orbitals of O, yielding  a metallic state, which is similar to the theoretical study of SrCrO$_3$.$^{28}$ The slight difference in the p orbitals of apical O  and in-plane O  reflects the tetragonal phase of crystal. The fact that the LSDA+U calculation still does not reveal the insulating nature of the real compound may simply be due to the electron correlations being too strong to properly capture the essential physics in a Hartree-Fock approximation. There may also be further effects not included in this calculation. For example, the Jahn-Teller (JT) effect has proven to be essential to understanding conductivity in some oxides. In LaMnO$_3$, even when using a more sophisticated  LDA+DMFT approach, the insulating behavior appears only when JT distortion and coulomb correction work together.$^{29}$ The ionic radii range from Ca$^{2+}$ to Ba$^{2+}$: r(Ca$^{2+}$(VIII)) = 1.12 \AA, r(Sr$^{2+}$(VIII)) = 1.26 \AA, r(Pb$^{2+}$(VIII)) = 1.29 \AA, r(Ba$^{2+}$(VIII)) = 1.42 \AA.$^{30}$ Correspondingly, the pseudocubic lattice parameter of the compound also increases, a$_c$(Ca) = 3.75 \AA, a$_c$(Sr) = 3.82 \AA, a$_c$(Pb) = 4.00 \AA, and a$_c$(Ba) = 4.07\AA. Due to the fully-filled orbital configurations of Ca$^{2+}$, Sr$^{2+}$, and Ba$^{2+}$, the electronic structures of these chromate compounds are mainly determined by the corner-sharing CrO$_6$ octahedra.  Therefore, one might think that the electronic phases of these chromate compounds, ranging from metallic to insulating, could be understood in a unified scheme, namely, a variation of the ionic radius modifies the octahedral unit cell, resulting in a Mott transition. However, the Mott transition is expected to be accompanied by a nonmagnetic to antiferromagnetic transition while similar materials, such as manganites, undergo a ferromagnetic to antiferromagnetic transition. In this case, the magnetism appears mostly unaffected by the conductivity change of the samples; an unexpected result. Thus the set of compounds CaCrO$_3$, SrCrO$_3$, and BaCrO$_3$ are a good model family for exploring more subtle aspects of a bandwidth controlled Mott transition especially the detailed connection between the magnetic structure and conductivity.
\vspace*{0.1cm}
 
   In conclusion, we report the epitaxial growth of a BaCrO$_3$ film with the perovskite structure, which to date has not been obtained in the powder samples made by standard solid-state synthesis techniques. This material, exhibiting weak ferromagnetism which is possibly due to a canted antiferromagnetic spin structure, has much in common with other chromate family members and should serve as an important example for understanding the magnetic and electronic structure of the TMOs. 

We thank L. Narangammana, H. E. Mohottala, and Y. F. Nie for helpful discussions. This work is supported by the NSF through grant DMR-0907197.


\scriptsize 
\begin{thebibliography}{99}

 \bibitem{run2det} M. Imada, A. Fujimori, and Y. Tokura, Rev. Mod. Phys. 70, 1039-1263 (1998).
 \bibitem{run2det} P. Schiffer, A. P. Ramirez, W. Bao, and S. -W. Cheong, Phys. Rev. Lett. 75, 3336 (1995).
 \bibitem{run2det}   D. I. Khomskii and G. A. Sawatzky, Solid State Commun. 102, 87 (1997).
 \bibitem{run2det}  Y. Yoshida, S. I. Ikeda, H. Matsuhata, N. Shirakawa, C. H. Lee and S. Katano, Phys. Rev. B 72, 054412 (2005).
 \bibitem{run2det}   B. L. Chamberland and C. W. Moeller, J. Solid State Chem. 5, 39 (1972).
 \bibitem{run2det}   B. L. Chamberland and C. W. Moeller, Solid State Commun. 5, 663 (1967).
 \bibitem{run2det}   J. B. Goodenough, J. M. Long, and J. A. Kafalas, Mater. Res. Bull. 3, 471 (1968).
 \bibitem{run2det}   J. F. Weiher, B. L. Chamberland, and J. L. Gillson, J. Solid State Chem. 5, 39 (1972).
 \bibitem{run2det}   W. L. Roth and R. C. DeVires, J. Appl. Phys. 38, 951 (1967).
 \bibitem{run2det}   J. S. Zhou, C. -Q. Jin, Y. -W. Long, L. -X. Yang, and J. B. Goodenough, Phys. Rev. Lett. 96, 046408 (2006).
 \bibitem{run2det}   A. J. Williams, A. Gillies, and J. P. Attfield, G. Heymann, H. Hupertz, M. J. Martinze-Lope, and J. A. Alonso, Phys. Rev. B 73, 104409 (2006). 
 \bibitem{run2det}   L. Ortega-San-Martin, A. J. Williams, J. Rodgers, J. P. Attfield, G. Heymann, and H. Huppertz, Phys. Rev. Lett. 99, 255701 (2007).
 \bibitem{run2det}  A. C. Komarek, S. V. Streltsov, M. Isobe, T. Moller, M. Hoelzel, A. Senyshyn, D. Trots, M. T. Fernandez-Diaz, T. Hansen, H. Gotou, T. Yagi, Y. Ueda, V. I. Anisimov, M. Gruninger, D. I. Khomskii, and M. Braden, Phys. Rev. Lett. 101, 167204 (2008).
 \bibitem{run2det}   S. V. Streltsov, M. A. Korotin, V. I. Anisimov, and D. I. Khomskii, Phys. Rev. B 78, 054425 (2008).
 \bibitem{run2det}   A. C. Komarek, T. Moller, M. Isobe, Y. Drees, H. Ulbrich, M. Azuma, M. T. Fernandez-Diaz, A. Senyshn, M. Hoelzel, G. Andre, Y. Ueda, M. Gruninger, and M. Braden, Phys. Rev. B 84, 125114 (2011).
 \bibitem{run2det}   P. A. Bhobe, A. Chainani, M. Taguchi, R. Eguchi, M. Matsunami, T. Ohtsuki, K. Ishizaka, M. Okawa, M. Oura, Y. Senba, H. Ohashi, M. Isobe, Y. Ueda, and S.Shin, Phys. Rev. B 83, 165132 (2011).
 \bibitem{run2det}   B. L. Chamberland, J. Solid State Chem. 43, 3 (1982).
 \bibitem{run2det}   P. S. Haradem, B. L. Chamberland,  L. Katz,  J. Solid State Chem. 1, 34 (1979).
 \bibitem{run2det}   C. -Q. Jin, J. S. Zhou, J. B. Goodenough, Q. Q. Liu, J. G. Zhao, L. X. Yang, Y. Yu, R. C. Yu, T. Katsura, A. Shatskiy, and E. Ito, Proc. Natl. Acad. Sci. USA, 105, 7115 (2008).

 \bibitem{run2det} G. Kresse and J. Furthmuller, Phys. Rev. B 54, 11169 (1996).

 \bibitem{run2det}G. Kresse and J. Furthmuller, Comput. Mater. Sci. 6, 15  (1996).

 \bibitem{run2det} G. Kresse and J. Hafner, J. Phys.: Condens. Matter 6, 8245 (1994).

 \bibitem{run2det}   Z. H. Zhu and X. H. Yan, J. Appl. Phys. 106, 023713 (2009).
 \bibitem{run2det}   O. Ofer, J. Sugiyama, M. Mansson, K. H. Chow, E. J. Ansaldo, J. H. Brewer, M. Isobe, and Y. Ueda, Phys. Rev. B     184405 (2010).
 \bibitem{run2det}   I. Dzyaloshinsky, J. Phys. Chem. Solids 4, 241 (1958).
 \bibitem{run2det}  T. Moriya, Phys. Rev. 120, 91 (1960).
 \bibitem{run2det} M. A. Korotin, V. I. Anisimov,  D. I. Khomskii and G. A. Sawatzky, Phys. Rev. Lett. 80, 4305 (1998).
 \bibitem{run2det} K. -W. Lee and W. E. Pickett, Phys. Rev. B 80, 125133 (2009).
 \bibitem{run2det} A. Yamasaki, M. Feldbacher, Y.-F. Yang, O. K. Andersen, and K. Held, Phys. Rev. Lett. 96, 166401 (2006).
 \bibitem{run2det}  R. D. Shannon, Acta Crystalloger. Sect. A 32, 751 (1976).

\end{thebibliography}
\end{document}